\documentclass{aa}

\usepackage{graphicx}
\usepackage{txfonts}

\begin{document}

\title{A Short Hard X-ray Flare from the Blazar NRAO 530 Observed by $INTEGRAL$\thanks{Based on 
observations obtained with \emph{INTEGRAL}, an ESA mission with instruments and science
data centre funded by ESA member states (especially the PI countries:
Denmark, France, Germany, Italy, Switzerland, Spain), Czech Republic and
Poland, and with the participation of Russia and the USA.}}

\author{L. Foschini\inst{1}, E. Pian\inst{2}, L. Maraschi\inst{3}, C.M. Raiteri\inst{4}, F. Tavecchio\inst{3}, 
G. Ghisellini\inst{3}, G. Tosti\inst{5}, G. Malaguti\inst{1}, G. Di Cocco\inst{1}}
\institute{INAF/IASF-Bologna, Via Gobetti 101, 40129 Bologna (Italy)
\and
INAF, Osservatorio Astronomico di Trieste, Via G.B. Tiepolo 11, 34131, Trieste (Italy)
\and
INAF, Osservatorio Astronomico di Brera, Via Bianchi 46, 23807, Merate (Italy)
\and
INAF, Osservatorio Astronomico di Torino, Via Osservatorio 20, 10025, Pino Torinese (Italy)
\and
Osservatorio Astronomico, Universit\`a di Perugia, Via B. Bonfigli, 06126 Perugia (Italy)
}

\offprints{\texttt{foschini@iasfbo.inaf.it}}
\date{Received 22 November 2005; Accepted 5 January 2006}

\abstract{We  report  about a  short  flare from  the blazar  NRAO  530
occurred on  $17$ February $2004$ and detected  serendipitously by the
IBIS/ISGRI  detector  on  board  \emph{INTEGRAL}. In  the  $20-40$~keV
energy  range, the  source,  that is  otherwise  below the  detection
limit, is detected at a level of $\approx 2\times 10^{-10}$~erg~cm$^{-2}$~s$^{-1}$
during a time interval of less than $2000$~s, which is about a factor 2 above 
the detection threshold.  At  other wavelengths, only nearly-simultaneous radio
data are available (1 observation  at $2$~cm on $11$ February $2004$),
indicating  a moderate  increase  of the  polarization. This  appears 
to  be the shortest  time variability episode ever detected in a high luminosity
blazar  at hard X-rays, unless the blazar is contaminated by
the presence of an unknown unresolved rapidly varying source.
}

%\titlerunning{}
\authorrunning{L. Foschini et al.}
 
\maketitle

\section{Introduction}
Blazars are, among active galactic nuclei (AGN), the most luminous and
most dramatically  variable.  These extreme properties  are likely due
to relativistic  aberration in a  kilo-parsec jet oriented at  a small
angle with  respect to our line  of sight, where plasma  moving with a
Lorentz  factor  of $\sim$10-20  causes  radiation  boosting and  time
foreshortening (see reviews by  Urry \& Padovani 1995,  Wagner \&
Witzel 1995, Ulrich et al. 1997).  In these sources, the maximum power
output  and  the  largest   variability  amplitudes  are  observed  at
high-energies,  from  X-   to  $\gamma-$rays,  therefore  blazars  are
interesting targets  for monitoring with  satellites like \emph{INTEGRAL}
(Pian  et  al.   1999).  During  its first  three  years  of  activity
\emph{INTEGRAL}  detected two blazars  in outburst,  with observations
triggered  by ground  telescopes  (S5 $0716+714$  Pian  et al.   2005,
3C$454.3$ Foschini et al.  2005, Pian et al., in preparation), and one
in  high state  (S5 $0836+710$),  serendipitously detected  during the
observation of  S5 $0716+714$ (Pian  et al. 2005).  Indeed,  thanks to
the large field-of-view of  the IBIS instrument onboard INTEGRAL, many
blazars, possibly in active  state, may be detected serendipitously by
the satellite  during its  long pointings.  In  order to  exploit this
advantage, we have  systematically searched the \emph{INTEGRAL} public
archive  for possible detection  of flaring  blazars.  We  report here
about the interesting detection  of a rapid flare apparently associated 
with the blazar NRAO~$530$.

NRAO~$530$ ($z=0.902$) is a  blazar belonging to the optically violent
variable  (OVV) quasar subclass,  with emission  lines in  the optical
spectrum, particularly H$\beta$ with  a rest-frame equivalent width of
$16\AA$, smaller  than  usually observed in OVVs  (Junkkarinen 1984; cf
also  Veron-Cetty \&  Veron 2000).   It exhibits  strong  outbursts in
radio (Bower  et al. 1997), optical  (up to $3$  magnitudes in $1977$,
Pollock  et  al. 1979,  Webb  et  al.  1988), X-ray (\emph{HEAO-1}, Marscher 
et al. 1979), and  $\gamma-$ray  bands (EGRET/CGRO, Mukherjee  et al. 1997), 
with  significant emission above $1$~GeV  (Lamb \&  Macomb  1997).  
The flare detected by the IBIS/ISGRI instrument on board \emph{INTEGRAL}
occurred  in February~$2004$ with a timescale ($\approx 2000$~s) much shorter than
any variability reported before.

In  the following,  we present  in detail the  \emph{INTEGRAL} data analysis and
results (Section 2),  data obtained quasisimultaneously in the radio (Section 3)
and a discussion of the relevance of the detected flare for blazars.

\begin{figure*}[!ht]
\centering
\includegraphics[scale=0.45,angle=270]{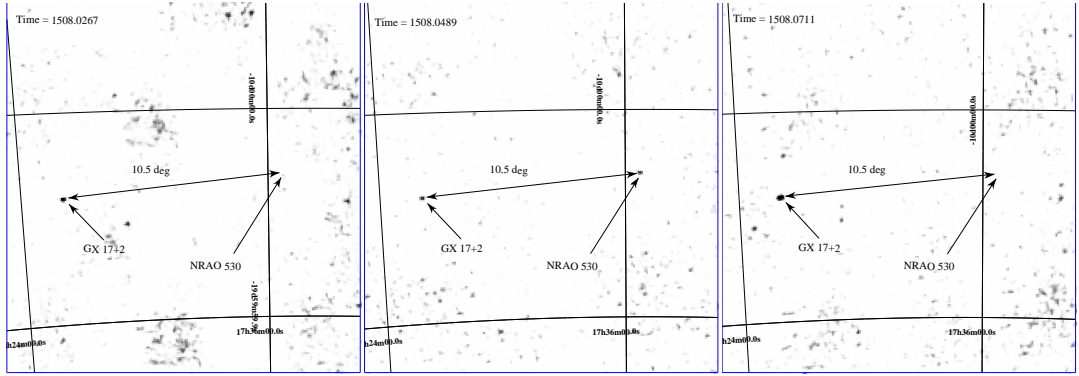}
\caption{IBIS/ISGRI significance map in the $20-40$~keV energy band for the pointing before, during,
and after the detection of NRAO $530$.}
\label{isgrimap}
\end{figure*}

\section{$INTEGRAL$ data analysis}
The \emph{INTEGRAL}  satellite (Winkler et  al. 2003) was  launched on
$17$  October $2002$  and  it  carries two  main  instruments for  the
$\gamma-$ray astrophysics, the imager IBIS ($0.02-10$~MeV, Ubertini et
al.   2003)  and  the  spectrometer  SPI  ($0.02-8$~MeV,  Vedrenne  et
al. 2003), plus two monitors, JEM-X ($3-35$~keV, Lund et al. 2003) and
OMC (V  filter, Mas-Hesse et  al. 2003).  Since NRAO~$530$  is located
close to the  Galactic Centre ($l=12^{\circ}.0$, $b=+10^{\circ}.8$), a
very crowded region, we do not use the data from the spectrometer SPI,
since it  has an angular  resolution of about $2^{\circ}.5$.  Also the
OMC  is strongly  limited by  the source  confusion:  its point-spread
function  (PSF) is $25''$  (FWHM).  The  US Naval  Observatory Catalog
(USNO B1, Monet et al.   2003) reports $12$ sources of brightness
comparable to  that of  NRAO~$530$ inside  a circular region  of $25''$
radius  centered around  the position  of the  blazar.  Thus,  the OMC
observations  are presumably  highly contaminated,  and  therefore not
used.

We also  checked the  JEM-X data for  possible detection,  however the
source was  never optimally located in the  monitor field-of-view: in only one
of the 11 pointings the source angular  distance from the FOV centre was
 $2^{\circ}.5$, less than the limit of $<3^{\circ}$ required for good
JEM-X performance.  Using the INTEGRAL pointing\footnote{Every single 
pointing is called ``Science Window'' (ScW) in the \emph{INTEGRAL} jargon. 
A ScW has a duration of about 2000 seconds.} where the source is detected
by IBIS (see below), which also corresponds to the 
minimum distance  from the JEM-X FOV center,  we  determine an  upper  
limit of  $30$~mCrab in  the $5-15$~keV energy  band ($5\sigma$ level, 
as  suggested in Westergaard et al. 2005).

Therefore, in the  following we will present only  the analysis of the
data acquired  by IBIS,  that is the  \emph{INTEGRAL} instrument
with  best sensitivity  in the  hard  X-ray range. The analysis has been
 performed using
the      Offline      Scientific      Analysis     (OSA)      software
package\footnote{Available  at \texttt{http://isdc.unige.ch/}.}, whose
algorithms for the reduction and treatment of the IBIS instrument data
are described in Goldwurm et al.~(2003).

The imager IBIS  (Ubertini et al. 2003) is  composed by two detectors,
ISGRI, sensitive to the radiation between $20$~keV and $1$~MeV (Lebrun
et  al.  2003),  and  PICsIT  ($175$~keV  to  $10$~MeV,  Di  Cocco  et
al.  2003), both coupled with the same tungsten  coded mask.  The latter  has the
pattern of  a modified uniformly  redundant array (MURA,  Gottesman \&
Fenimore  1989), with a  basic configuration  of $53\times  53$ square
pixels $11.2$~mm sized.  The mask is separated from the ISGRI layer by
about $3200$~mm, so that the  angular distance of ghosts in the system
point spread function (SPSF) of ISGRI is about $10^{\circ}.5$ (Gros et
al. 2003).  This means that sources with  reciprocal angular distances
of that size  can be confused and specific care  is necessary in order
to disentangle the different contributions.

On  February 17, $2004$ (orbit $164$), while \emph{INTEGRAL}  was  
executing  the  Galactic  Centre  Deep  Exposure (GCDE), ISGRI detected 
a signal during the time interval between $00^{\rm h}:40^{\rm m}:05^{\rm s}$ 
and $01^{\rm h}:09^{\rm m}:18^{\rm s}$ UTC at $\alpha=17^{\rm h}:33^{\rm m}:04^{\rm s}$ 
and $\delta=-13^{\rm h}:02^{\rm m}:50^{\rm s}$ (J2000). 
The $3'-$radius uncertainty region  associated with this centroid is consistent 
with  the blazar NRAO~$530$ ($\alpha=17^{\rm h}33^{\rm m}02^{\rm s}.71$,
$\delta=-13^{\circ}04'49''.5$, J2000) and one ghost of the SPSF of the neutron 
star GX~$17+2$ ($\alpha=18^{\rm   h}16^{\rm m}01^{\rm s}.40$,  $\delta=-14^{\circ}02'11''.0$,
J2000). The angular distance  between the  two sources  is precisely
$10^{\circ}.5$,  so  that the  two  SPSF  overlap (Fig.~\ref{isgrimap}). 
However, one  can disentangle  the contribution of  the two  sources by  
analysing their time evolution: if one SPSF is contaminated by the other, 
both sources should show the same time behaviour.

\begin{figure*}[!ht]
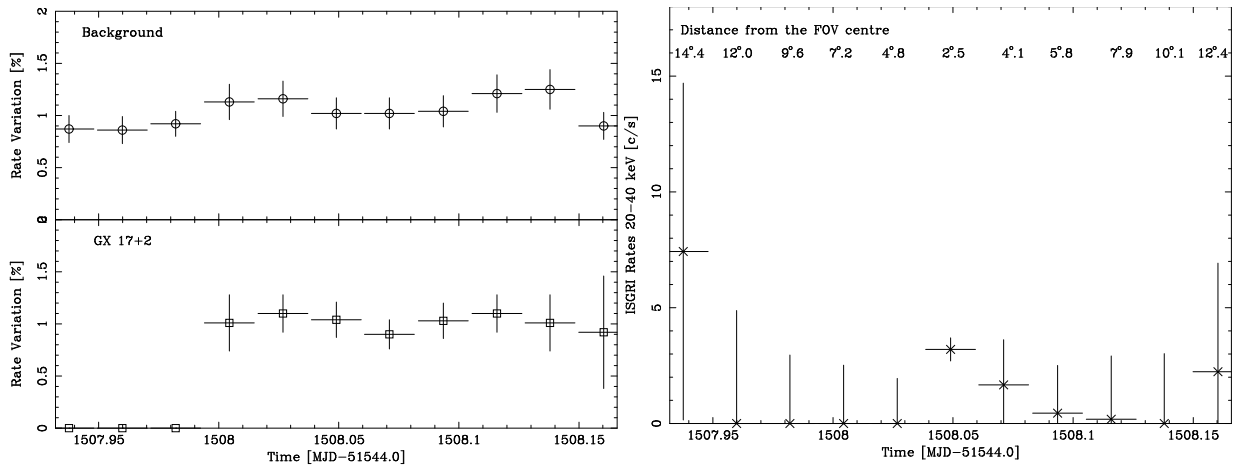

\centering
\includegraphics[angle=270,scale=0.35]{4804_f2.ps}
\includegraphics[angle=270,scale=0.35,clip,trim=18 0 0 0]{4804_f3.ps}
\caption{(\emph{left,top}): Background rates normalized to the weighted average ($79\pm 12$~c/s). (\emph{left,bottom}): 
GX~$17+2$ rates normalized to the weighted average ($10.9\pm 0.9$~c/s); in the first 3 ScW the source was outside the 
ISGRI FOV. (\emph{right}): NRAO~$530$ rates; upper limits are at $3\sigma$. In the top of the figure, the distance of
the source from the ISGRI FOV centre is shown in every ScW. All the lightcurves are in the $20-40$~keV 
energy band.}
\label{fig1}
\end{figure*}

Since the  latest version of  \texttt{OSA 5.0} for  IBIS/ISGRI removes
the  sources located  at  the  potential positions  of  the ghosts  of
brightest sources, we used  an older version (\texttt{OSA 3.0}), where
this  feature was  not  yet implemented.   We  also cross-checked  the
consistency  of  the  results  obtained  with  \texttt{OSA  3.0}  with
\texttt{OSA 5.0},  however, in  the latter case,  we adopted  an input
source catalog containing only NRAO~$530$  as a source, to by-pass the
above  protection.  Obviously,  we also  checked that  the  results on
GX~$17+2$ were not affected when adopting this solution, by performing
a run with \texttt{OSA 5.0} and the full input catalog.

We compared  the behaviour of the  background and of the  2 sources by
analyzing the 5 pointings before and after the event, corresponding to
the time  interval between February $16^{\rm th}$,  $2004$ at $21^{\rm
h}:59^{\rm  m}:08^{\rm  s}$, and  February  $17^{\rm  th}$, $2004$  at
$03^{\rm h}:49^{\rm m}:44^{\rm s}$ UTC.  We normalized the count rates
of the background  and of GX~$17+2$ in the  $20-40$~keV energy band to
their weighted averages. For NRAO~$530$, which has one detection only,
we  left  the  count   rates  unchanged.   Fig.~\ref{fig1}  shows  the
lightcurves  of  the  background,  GX~$17+2$, and  NRAO~$530$.   While
GX~$17+2$ follows  the background time evolution  (Fig.~\ref{fig1}, left), this
is not  the case  for NRAO~$530$ (Fig.~\ref{fig1},  right).  Were  the putative
detection  of NRAO~$530$  only a  ghost  of GX~$17+2$,  then it  would
follow the same temporal behaviour, which would contradict the derived
 upper limits (see also Fig.~\ref{isgrimap}).

No signal was  detected in any other pointing  at that position during
the  GCDE, nor  during  any  other INTEGRAL  observation  of this  sky
region.  The  difference in the  upper limits on the  NRAO~$530$ count
rate in the various pointings  (Fig.~\ref{fig1}, right) is due to the different
location of NRAO~$530$ with respect  to the ISGRI FOV: the farther the
source from the  FOV center, the more shallow its  upper limit, due to the
radially decreasing  sensitivity of the  ISGRI detector.  Note  that the
upper limits obtained in the $\sim$2 hours bracketing the NRAO  530
detection indicate that the flux change is highly significant.  Therefore,
the hypothesis that the source was
constant and  could only  be detected when  sufficiently close  to the
ISGRI FOV center is ruled out.   We conclude that the detection of the
rapid variation is genuine. 

The total  time on source is $1745$~s, that can be considered
the minimum time scale of the event. It is worth mentioning that \emph{INTEGRAL}
performs two slews, one before and the other after the pointing, each $\approx 120$~s
long. It is not possible to detect the source during a slew in a 
coded-mask instrument (except for very bright cases, like GRB) and therefore
it is not possible to say if NRAO~$530$ was flaring already during the slews.
Taking into account also the marginal detection in the pointing following the central
one the maximum time scale of the event is $\approx 1 $~h.

We searched the SIMBAD catalog for Galactic  hard X-ray  sources within  
a $6'$  radius (twice  the ISGRI error  radius  for  a  source  with this  
significance,  see  Gros  et al.  2003)  of  the   position  of  NRAO~$530$,  
that  could  possibly contaminate  our  ISGRI  detection,  but  found  none.   
Therefore,  we attribute the flux detection entirely to NRAO~$530$.

The count rates of NRAO~$530$  in the $20-40$~keV energy band, already
corrected   for  systematics,   are:  $2.7\pm   0.5$~c/s   when  using
\texttt{OSA 3}, that correspond to $26\pm 5$~mCrab\footnote{We assumed
$1~{\rm  Crab}  = 105$~c/s  in  the  $20-40$~keV  energy band  of  the
IBIS/ISGRI detector.}. When using  \texttt{OSA 5}, the rate is $3.2\pm
0.6$~c/s,  corresponding  to  $30\pm  6$~mCrab,  consistent  with  the
measurement  with  \texttt{OSA  3}.  We  take a  weighted  average  as
reference: $28\pm  3$~mCrab. By adopting the standard  spectrum of the
Crab  as reported  by Toor  \& Seward~(1974)\footnote{Power  Law model
with             $\Gamma=2.1$             and            normalization
$9.7$~ph~cm$^{-2}$~s$^{-1}$~keV$^{-1}$  at  $1$~keV.},  the  converted
flux is $(2.1\pm 0.2)\times 10^{-10}$~erg~cm$^{-2}$~s$^{-1}$. No detection
has been found in PICsIT, the high-energy layer of IBIS.

\section{Radio data}
We searched  also for quasi-simultaneous  data at
other wavelengths and found only a radio observation at $2$~cm (VLBA),
made on  $11$ February $2004$,  a few days before  the \emph{INTEGRAL}
detection,    within   the    MOJAVE
project\footnote{\texttt{http://www.physics.purdue.edu/astro/MOJAVE/}}
(Lister \& Homan  2005).  The flux density was  $4.1$~Jy, with $3.1$\%
linear  polarization.  The  previous  observation dates  back  to  $9$
October $2002$,  with a higher flux ($5.7$~Jy)  but lower polarization
($0.5$\%).  In 2005, lower  fluxes and  linear polarization  have been
observed.  These  data  can  be  compared with  the  historical  radio
observations by Bower et al. (1997): as noted already by those authors
there is  no apparent correlation  between the radio flux  density and
the optical activity in $1977-1978$. NRAO~$530$ generally shows fluxes
from  a  few  up  to   more  than  $10$~Jy,  depending  on  the  radio
frequency. Therefore,  the MOJAVE  2004 measurement of  $4.1$~Jy could
indicate a low activity phase.

\begin{figure}[!t]
\centering
\includegraphics[scale=0.45]{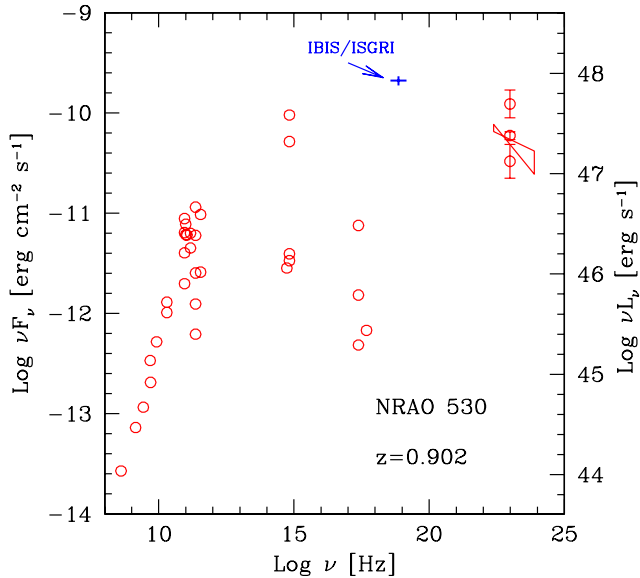}
\caption{Spectral energy distribution of NRAO~$530$ updated with the \emph{INTEGRAL} data of the present
work. Radio data from NED and Bower et al. (1997). Reference value of optical observations 
from Hewitt \& Burbidge (1993); historical maximum from Pollock et al. (1979), 
Webb et al. (1988). X-rays: \emph{ROSAT} (Comastri et al. 1997), \emph{Einstein} 
(Marscher \& Broderick 1981), \emph{HEAO-1} (Marscher \& Broderick 1981).
$\gamma-$rays: \emph{CGRO}/EGRET (Hartman et al. 1999). Blue cross:
IBIS/ISGRI, present work.}
\label{fig2}
\end{figure}

The polarization measurements reported by  Bower et al. (1997) show an
increase (up to about $8$\% at $8$~GHz) during the two years preceding
the  optical  outburst  in  $1977-1978$.   In  our  case,  the  MOJAVE
measurement of  a moderately high polarization value  some days before
the hard-X flare  could be linked  to the latter. We note that the present  
episode is a rapid event, not a long outburst  like  in $1977-1978$,  
therefore the fact that only moderate polarization was observed is not inconsistent
with the fast flare in hard X-rays.

\section{Discussion}
Hard X-ray emission from the blazar NRAO~$530$ was detected 
serendipitously by \emph{INTEGRAL} with IBIS/ISGRI during a long exposure 
of the Galactic Center. During a time interval lasting $\approx 2000$~s 
the source flux in the $20-40$ keV range rose above the detection threshold 
by about a factor of $\sim 2$ and then faded again.

The  spectral  energy  distribution  of  NRAO~$530$  constructed  with
historical data and with  the present \emph{INTEGRAL} flux is reported
in Fig.~\ref{fig2}. Interestingly, although very high, the IBIS/ISGRI point
is consistent with the extrapolation to lower energies of the average spectrum 
observed in the EGRET energy range.
The variability amplitude measured by EGRET was a factor $6$ and rather erratic, 
still  on  timescales of weeks (Mukherjee et al. 1997). On the other hand, timescales 
as short as implied here would not have been accessible for EGRET due to the low counting rate. 
The hard X-ray flux measured during the flare for any reasonable spectral shape would 
imply a much larger X-ray flux than observed previously. Large variations have been 
observed from this object also in the optical (more than a factor 10 over timescales 
of years), thus overall the amplitude of variability is not unprecedented.

What is truly exceptional in this event is the short timescale, less than
an hour in an object of extremely high luminosity. This may cast doubts 
on the association with NRAO~$530$. The event could then be attributed
to some still unknown galactic source inside the $3'$ radius circle
of positional uncertainty in the ISGRI detection. In this case, only future
satellites, like SIMBOL-X (Ferrando et al. 2005), will have the necessary 
spatial resolution at hard X-rays to confirm the association with NRAO~$530$.
For the moment, with the present data, instrument performances, and catalogs, 
it is interesting to consider the possibility that this highly significant variation 
is associated with the blazar.

Rapid flares, with timescales of thousands of seconds in the X-rays,
are often observed in BL Lac sources, especially in the TeV sources
(e.g. Brinkmann et al. 2005), but they are rarely seen in
blazars belonging to the quasar class, where typical timescales, even
in hard X-ray--$\gamma$-ray band, are in the range of several
hours--days (e.g. Wagner \& Witzel 1995). It is worth noting that FSRQ are
generally faint in X-rays and therefore the above assertion could be biased in
the past by the lack of high-energy astrophysics satellites with the
necessary sensitivity. The present and the next generation of X- and
$\gamma-$ray satellites should remove this bias.

The interpretation of such a rapid flare in the context of the
standard synchrotron-inverse Compton models for blazars
(e.g. Ghisellini et al. 1998) requires quite extreme conditions for
the emitting source. A flare lasting less than $\Delta t=3600$~s must be
emitted by a region with a size $R < 1.1\times 10^{14} \delta$ cm, where
$\delta$ is the Doppler factor. Adopting the value of $\delta=15$, suggested by VLBI 
observations (Bower \& Backer 1998), the size of the source is
$R< 1.6 \times 10^{15}$~cm, an order of magnitude less than dimensions
typically estimated for blazars ($10^{16}-10^{17}$~cm). In the case of 
a mass of the central supermassive black hole of $10^{8}M_{\odot}$,
the above distance would correspond to only about $50$ times the radius of the 
innermost stable orbit.

A direct alternative is to admit a large value of the Doppler factor, 
$\delta \sim 100$. Such large values of $\delta$ are sometimes invoked to 
explain the rapid intra day variability observed in the radio band 
(e.g. Wagner \& Witzel 1995). In this case, it would be possible to explain
the short flare in terms of unsteadyness of the jet flow, due to a single non
stationary shock (Hughes et al. 1985) possibly induced by a collision of 
two relativistic plasma shells in the jet (internal shock, Spada et al. 2001). 
The event discussed here could also result from an internal shock developping 
very close to the origin of the jet with less extreme values of $\delta$.
The moderate polarization observed at radio wavelengths is consistent with these
scenarios, since it arises from regions much further out in the jet.

Although the data available are few and sparse,  it is possible
that the present observation represents the ``tip of an iceberg'' leading the way to
discover more short time scale events. We plan to continue our search of variability
of blazars in the  \emph{INTEGRAL} field of view.

\begin{acknowledgements}
LF wishes to thank A.~Domingo Garau for useful discussion about OMC data.
This research has made use of the SIMBAD database, operated at CDS, Strasbourg, France,
of the NASA/IPAC Extragalactic Database (NED) which is operated by the 
Jet Propulsion Laboratory, California Institute of Technology, under contract with the National 
Aeronautics and Space Administration, of data obtained from the High Energy Astrophysics Science Archive 
Research Center (HEASARC), provided by NASA's Goddard Space Flight Center, and of data from
the Monitoring of Jets in AGN with VLBA Experiments (MOJAVE) Project.
This work was partly supported by the European Community's Human Potential
Programme under contract HPRN-CT-2002-00321 (ENIGMA).
\end{acknowledgements}

\vskip 12pt
\emph{Note added in proof}. After acceptance of the manuscript, the
Swift satellite performed two observations of NRAO530 (on 10
and 14 February 2006). During both observations, the X-Ray
Telescope (XRT) detected only the blazar NRAO 530 within
the ISGRI error circle of $3'$. This result strengthens our conclusion
that the exceptional flare detected by INTEGRAL is very
likely due to NRAO 530.

\end{document}